\documentclass[preprint,nofootinbib]{revtex4}
\usepackage{epsfig}
\usepackage{amssymb}
\newcommand{\CC}{\Lambda}
\newcommand{\rL}{\rho_{\CC}}
\newcommand{\rLo}{\rho_{\CC 0}}

\newcommand{\rLb}{\rho_{\CC}^{({\rm b})}}

\newcommand{\rVu}{\rho_{\rm vac}^{(1)}}
\newcommand{\Gb}{G^{({\rm b})}}

\newcommand{\aeq}{a_{\rm eq}}
\newcommand{\ah}{\hat{a}}
\newcommand{\Tgz}{T_{\gamma 0}}

\begin{document}
\begin{titlepage}

\begin{center}
\small{This essay received an \textit{Honorable Mention}
from the Gravity Research Foundation (2015)\\ Awards Essays on
Gravitation}
\end{center}
\vspace{1cm}

\vspace{0.5cm}

\begin{center}
{\large \bf The cosmological constant and entropy problems:
mysteries of the present with profound roots in the past}

\vskip 0.5cm {\large \bf Joan Sol\`a}
\end{center}

\begin{quote}
\begin{center}
High Energy Physics Group, Departament ECM, \\ and Institut de
Ci\`encies del Cosmos, Universitat de Barcelona,\\ Av.
Diagonal 647 E-08028 Barcelona, Catalonia, Spain\\
Email: sola@ecm.ub.edu
\end{center}
\end{quote}
\vspace{0.2cm} \centerline{(Submission date: March 30, 2015)} \vspace{0.2cm}
\centerline{\bf Abstract}
\bigskip
An accelerated universe should naturally have a vacuum energy
density determined by its dynamical curvature. The cosmological
constant is most likely a temporary description of a  dynamical
variable that has been drastically evolving from the early
inflationary era to the present. In this Essay we propose a unified
picture of the cosmic history implementing such an idea, in which the
cosmological constant problem is fixed at early times. All the main
stages, from inflation and its  (``graceful'') exit into a standard
radiation regime, as well as the matter and dark energy epochs, are
accounted for. Finally, we show that for a generic Grand Unified
Theory associated to the inflationary phase, the amount of entropy
generated from primeval vacuum decay can explain the huge measured value
today.

\vspace{0.3cm}


\end{titlepage}

\pagestyle{plain} \baselineskip 0.75cm

\section{The cosmological constant problem}

One of the most perplexing aspects of the cosmological constant (CC)
or $\CC$-problem\,\cite{Weinberg89,CCP1,CCP2,CCP3} is that the
current  energy density $\rLo=\CC/(8\pi G)$ of vacuum is of order
$10^{-47}$ GeV$^4\sim \left(10^{-3} {\rm eV}\right)^4$ in natural
units ($G$ is Newton's constant). Only a light neutrino of a
millielectronvolt could be associated to such low energy density.
Any other known particle provides a contribution which is
exceedingly much bigger, e.g. $m_e^4/\rLo\sim 10^{34}$ for the
electron.

With the advent of the Higgs boson discovery, the existence of the
electroweak vacuum is generally considered a proven fact. Denoting
by $M_H\simeq 125$ CeV the (measured) Higgs mass and $M_F\equiv
G_F^{-1/2}\simeq 293$ GeV the Fermi scale, the zero-point energy
(ZPE) from the Higgs field is of order $M_H^4\sim 10^8$ GeV$^4$, and
the ground state value of the (classical)  Higgs potential reads
$\langle V\rangle=-(1/8\sqrt2) M_H^2\,M_F^2\sim -10^9$ GeV$^4$. In
magnitude this is of order $\sim v^4$, where $v\sim 250$ GeV is the
vacuum expectation value of the scalar field. Equally significant is
the ZPE from the top quark (with mass $m_t\simeq 174$ GeV), which is
of order  $m_t^4\sim 10^9$ GeV$^4$ and negative (because it is a
fermion). After adding up all these effects a result of the same
order ensues which is $\left(10^9/10^{-47}\right)\sim 10^{56}$ times
bigger than what is needed. Even if by some miracle they would
conspire to give zero at the tree-level, higher order effects (both
from the ZPE and the Higgs potential) would spoil the cancelation
until $\sim20$th order of perturbation theory -- still rendering a
quantum payoff of order $\rLo$:
\begin{equation}\label{nloops}
\delta\rho_{\CC}\sim\left(\frac{g^2}{16\,\pi^2}\right)^{20}\,
M^4\sim 10^{-47}\,GeV^4\,.
\end{equation}
Here $g$ stands typically for the $SU(2)$ electroweak gauge
coupling, and $M$ is an effective mass of order  $\sim v$.
Accordingly, the very many thousand electroweak loops contributing
all the way down from 20th order of perturbation theory to zeroth
order should be carefully, and somehow magically, adjusted. Truly
bewildering, if formulated on these grounds!

\section{Renormalizing away $\CC$}

What could possibly be wrong in the above argument? Most likely
this: to accept uncritically that the $\sim m^4$ contribution from a
particle of mass $m$, and the $\sim v^4$ one from the Higgs
potential, are both individual {\em physical} contributions to
$\rL$. Let us have a closer look at the origin of the $\sim m^4$
effects by considering a real scalar particle in the context of
quantum field theory in curved spacetime\,\cite{BirrellDavis}. The
one-loop correction to the effective action can be computed e.g. in
the $\overline{\rm MS}$ scheme in $N$ dimensions. After setting
$N\to 4$, except in the poles, one arrives at:
\begin{eqnarray}\label{eq:Gammar1comp}
\Gamma^{(1)}_{\rm eff}= \frac{1}{32\pi^2}\int {\rm d}^4 x \sqrt{-g}
\left(\frac{2}{4-N}+\ln\frac{\mu^2}{m^2}+C\right)
\left(\frac12\,m^4-m^2\,a_1(x)+a_2(x)+\cdots \right)\,,
\end{eqnarray}
where $C$ is a constant. The $a_{1,2}$ are the so-called
Schwinger-DeWitt coefficients (coming from the adiabatic expansion
of the matter field propagator in the curved background). They have
a purely geometric form given by a linear combination of the
curvature $R$ and the higher derivative (HD) terms $R^2$,
$R_{\mu\nu}^2$ etc. If the starting classical action contains
already the usual Einstein-Hilbert (EH) term and the HD terms, all
divergences can be absorbed by the counterterms generated from the
bare parameters $\rLb$, inverse Newton's coupling and the
coefficients $\alpha_i^{\rm (b)}$ of the various higher order $\sim
R^2$:
\begin{equation}\label{eq:split2}
\rLb=\rL(\mu)+\delta\rL\,,\ \ \
\frac{1}{\Gb}=\frac{1}{G(\mu)}+\delta\left(\frac{1}{G}\right)\,,\ \
\ \alpha_i^{\rm (b)}=\alpha_i(\mu)+\delta\alpha_i\,.
\end{equation}
As usual, each bare quantity splits into the renormalized part
(carrying an arbitrary scale $\mu$) and a counterterm, which is then
chosen to cancel the corresponding divergence (pole at $N=4$).
Gathering the various pieces that go into the renormalization of the
vacuum energy density, we are led to the expression:
\begin{equation}\label{deltaMSBPhysical}
\rL(\mu)+\delta\rL-\frac{m^4}{64\pi^2}\,\,\left(\frac{2}{4-N}
+\ln\frac{\mu^2}{m^2}+C\right)\,.
\end{equation}
In the $\overline{\rm MS}$ scheme the counterterm reads
${\delta}\rL^{\overline{\rm MS}}=\left({m^4}/{64
\pi^2}\right)\,\left(\frac{2}{4-N}+{\rm const}.\right)$ and the
one-loop renormalized result takes on the form:
\begin{equation}\label{renormZPEoneloopCurved}
\rVu=\rL(\mu)+\frac{m^4}{64\pi^2}\,\left(\ln\frac{m^2}{\mu^2}+C_{\rm
vac} \right)\,,
\end{equation}
where $C_{\rm vac}$ is a finite constant. We remark that the very
same result (\ref{renormZPEoneloopCurved}) is obtained from the
(much) simpler calculation of the ZPE in flat spacetime, starting
e.g. from the dimensionally regularized sum of half frequencies
$\sum_k\frac12 \omega_k$ in the continuum limit. The difference is
that, in the curved spacetime treatment (\ref{eq:Gammar1comp}), the
geometric terms also appear and involve corrections to the EH action
(hence to Newton's coupling) and the HD terms $\sim R^2$.

From Eq.\,(\ref{renormZPEoneloopCurved}) one usually concludes that
a particle of mass $m$ contributes a quantum correction $\sim m^4$
to the value of the CC; this is what triggers the preposterous fine
tuning problem outlined in the beginning.  However, we can take
another viewpoint/ansatz.

Recall that the counterterm depends on an arbitrary constant and
that the renormalized $\rL(\mu)$ is, albeit finite, {\emph not} a
physical quantity.  Thus, being the formal expression
(\ref{renormZPEoneloopCurved}) the same in flat spacetime, a more
natural  renormalization condition is to arrange for the exact
cancelation of both the UV \emph{and} the finite parts with the
counterterm. In other words, we should set the expression
(\ref{deltaMSBPhysical}) exactly to zero. In this way $\rVu=0$ in
flat spacetime rather than the result (\ref{renormZPEoneloopCurved})
-- incompatible with Einstein's equations in that background. By the
same token the Higgs yield $\sim v^4$, which is part of the flat
space result, is included in that prescription. Overall the CC is
zero at this point and flat spacetime can be a solution of the field
equations in the absence of expansion.  There is no fine tuning now:
for we did not adjust finite numbers; rather, we renormalized an
infinite quantity carrying an arbitrary finite part, similar to what
is done e.g. with the mass and charge of the electron in QED. Furthermore, the renormalization is meaningful as it is carried out in the UV regime, where we have the reliable tools of QFT to effectively implement the renormalization program.

\section{RUNNING VACUUM}

Removing the flat spacetime result (\ref{renormZPEoneloopCurved}) is
similar to subtracting the free space part in the Casimir effect so
as to project the vacuum vibrational modes in between the plates
only. In our case what remains, after renormalizing away the
unwanted terms in the CC, are just the EH and HD curvature effects.
But we expect something else, to wit: the genuine vacuum
contributions related to the expanding background. They should also
be purely geometric and of quantum nature, for only in this way can
they be as small as are currently observed. The leading effects may
generically be captured from a renormalization group equation of the
form
\begin{eqnarray}\label{seriesRLH}
\frac{d\rho_{\Lambda}}{d\ln
H^2}=\frac{1}{(4\pi)^2}\sum_{i}\left[\,a_{i}M_{i}^{2}\,H^{2}
+\,b_{i}\,H^{4}+c_{i}\frac{H^{6}}{M_{i}^{2}}\,+...\right] \,.
\end{eqnarray}
This equation describes the rate of change of $\rho_{\Lambda}$ with
the Hubble function $H(t)$, acting here as the running scale in the
FLRW metric -- see \cite{JSPRev2013} for a review and a comprehensive list of references. The r.h.s. of (\ref{seriesRLH}) represents the
$\beta$-function of $\rL$. It can involve \emph{only} even powers of
the Hubble rate $H$ (because of the covariance of the effective
action)\,\cite{ShapSol02,Fossil07}. The coefficients $a_i$,
$b_i$,$c_i$... are dimensionless, and the $M_i$ are the masses of
the particles in the loops. For a concrete scenario
of this kind within anomaly-induced inflation, see \,\cite{Fossil07}; and for an implementation within dynamically broken Supergravity cf. \,\cite{BasMavroSol2015}. Related developments within the renormalization group approach in cosmology are presented in \cite{Markkanen2015}.

Integrating the above equation and keeping, for the sake of
illustration, only one of the higher powers
$H^{n+2},$ we can express the result as follows:
\begin{equation}\label{LambdaH2Hn}
\rL(H) = \frac{3M_P^2}{8\pi}\left[c_0 + \nu H^{2} +
\frac{\phantom{...}H^{n+2}}{H_{I}^{n}}\right] \ \ \ \ \ \ \
(n\geq2)\,,
\end{equation}
with $c_0$ an integration constant and $H_{I}$ a parameter, both
dimensionful\,\cite{LBS2013}. For a generic GUT, the dimensionless parameter
$\nu=\frac{1}{6\pi}\, \sum_{i=f,b} a_i\frac{M_i^2}{M_P^2}$ is feeded
by the heavy masses $M_i$ of boson and fermions relative to the
Planck mass $M_P$. Typically
$|\nu|=10^{-6}-10^{-3}$\,\cite{Fossil07}.

\section{Inflation and graceful exit}

We adopt (\ref{LambdaH2Hn}) as a prototype for a ``unified running
vacuum model'' in the framework of the gravitational field equations
within the FLRW metric in flat $3$-dimensional space:
\begin{equation}\label{eq> generalizedFriedmann}
3H^2=8\pi
G(\rho_m+\rL(H))\,,\ \ \ \ \ \ \ \  2\dot{H}+3H^2=-8\pi
G(\omega_m\rho_m-\rL(H)).
\end{equation}
Taking  $\omega_m=1/3$  for the equation
of state parameter of the relativistic matter fluid in the early
universe, and neglecting $\nu$ and $c_0/H^2$ at this stage, we can
solve for $H$ and the energy densities in terms of the scale factor.
We find:
\begin{equation}\label{eq:Ha}
H(\ah)=\frac{H_I}{\left(1+\ah^{2n}\right)^{1/n}}
\end{equation}
along with
\begin{equation}\label{eq:densities}
\rho_\Lambda(\ah)=\frac{\rho_I}{f(\ah)}\,,\ \ \ \ \ \ \rho_r(\ah)=\frac{\rho_I\ah^{2n}}{f(\ah)}\,,
\end{equation}
where
\begin{equation}\label{eq:FAHAT}
f(\ah)\equiv\left(1+\ah^{2\,n}\right)^{1+2/n}\,.
\end{equation}
Here $\ah=a/\aeq$ is the scale factor, normalized to the value $\aeq$ where the decaying vacuum density equals the radiation density (i.e.
$\rho_{\CC}=\rho_r$ at $\ah=1$);  and $\rho_I=3H_I^2/8\pi G$ is the
(finite) critical density at $a=0$.

At the initial point of the cosmic evolution, $\rho_{\CC}(0)=\rho_I$
and $\rho_r(0)=0$, so the model is nonsingular. Furthermore, for a
typical GUT scale $M_X\sim 10^{16}$ GeV  associated to the
inflationary epoch, we find $H(\ah)<H_I\sim\sqrt{\rho_I}/M_P\sim
(M_X/M_P)^2M_P<10^{-5}M_P$ since $\rho_I\sim M_X^4$. This result is
in agreement with the well known CMB bound on the fluctuations
induced by the tensor modes\,\cite{LiddleLyth}.

Let us briefly highlight the expansion and thermal histories. In the
beginning we have $\ah\to 0$ and  $f(\ah)\to 1$, and hence  $\rL\simeq\rho_I=const.$ and $\rho_r\to 0$. Thus the universe starts with no matter at all; it contains just vacuum energy and as a result grows
exponentially fast: $a(t)\propto {\rm e}^{H_{I}t}$. At the same time
$\rho_r(\ah)$, which starts from zero, increases very fast as $\sim \ah^{2n}$ at the expense of vacuum
decay. Much later (when $\ah\gg1$, $f(\ah)\to \ah^{2n+4}$) we attain the asymptotic regime within the radiation epoch, in which the relativistic matter density decays as
$\rho_r\sim a^{-4}$ ($a\sim t^{1/2}$). Thus we achieve ``graceful
exit'' from inflation into the standard radiation epoch, a
remarkable feature which -- we should emphasize -- holds good for
\emph{any} $n$ in (\ref{LambdaH2Hn}).

\section{Solving the cosmological entropy problem}

The temperature of the heat bath generated from primeval vacuum
decay follows from equating $\rho_r(a)$ to the black-body form
$(\pi^2/30) g_{*} T^4_r$, where $g_{*}$ is the number of active
d.o.f. Thus  $T_r(\ah)=T_X \ah^{n/2}/f^{1/4}(\ah)$, where
$T_X\equiv\left({30 \rho_I}/\pi^2g_{*}\right)^{1/4}\sim M_X$ is of
the order of the maximum attained temperature. The radiation entropy
$S_{r}= \left(4\rho_r/3T_r\right) a^3$  \,\cite{KolbTurner} now
yields:
\begin{equation}\label{eq:entropy}
S_{r}(\ah)= \left(\frac{4\rho_I}{3T_X}\right)\,g(\ah)\aeq^3\,, \ \ \
\ \ \
g(\ah)\equiv\frac{\ah^{3(1+n/2)}}{\left[1+\ah^{2n}\right]^{\frac34(1+2/n)}}\,.
\end{equation}
It rises extremely fast in the beginning: $S\sim \ah^{3(1+n/2)}$
(e.g $S\sim \ah^6$ for $n=2$). But deep in the radiation epoch
$\ah\gg1$ (i.e. for $a\to a_r\gg\aeq$) we have $g(\ah)\to 1$, and
$S_r$ rapidly stagnates at the asymptotic value $S_r\to
S_{\infty}\equiv\left(4\rho_I/3T_X\right)
\aeq^3=(2\pi^2/45)g_{*}T_X^3\aeq^3$. Upon inspecting once more the temperature
$T_r(\ah)$,  we find that in the beginning it also rises fast:
$T_r\sim \ah^{n/2}$, but for $a\to a_r$ it eventually adapts to the
adiabatic behavior  $T_r=T_X/\ah_r$ (for \emph{any} $n$), i.e.
\begin{equation}\label{eq:adiabatic}
T_X\aeq=T_r a_r\,,\ \ \ \ \ \ \ (\ah\gg 1)\,,
\end{equation}
thereby the asymptotic entropy can be cast as
\begin{equation}\label{eq:Sinfinite}
S_{\infty}=(2\pi^2/45)g_{*}T_r^3 a_r^3\,.
\end{equation}
It is precisely during
this adiabatic phase  when the quantity $g_{*}T_r^3 a_r^3$ becomes
conserved and  equals the current value
$g_{s,0}\,\Tgz^{3}\,a_0^{3}$, in which $\Tgz\simeq 2.725$$^{\circ}$K
(CMB temperature now) and  $g_{s,0}=2+6\times
(7/8)\left(T_{\nu,0}/\Tgz\right)^3\simeq 3.91$ is the entropy factor
for the light d.o.f. today, computed from the ratio of the present
neutrino and photon temperatures. The upshot is that the entropy
enclosed in our horizon today, $H_0^{-1}$, namely
\begin{equation}\label{eq:S0}
S_{0}=
\frac{2\pi^2}{45}\,g_{s,0}\,\Tgz^3\,\left(H_0^{-1}\right)^3\simeq
2.3 h^{-3} 10^{87}\sim 10^{88} \ \ \ \ \ \ (h\simeq 0.67),
\end{equation}
can be fully accounted for from the asymptotic value $S_{\infty}$ in
the radiation epoch. Such number, therefore, was deeply encoded in
our remote past and -- remarkably enough -- does not depend neither
on the details of the GUT nor on the value of $n$ in
Eq.\,(\ref{LambdaH2Hn}). This result generalizes the one found for $n=2$ in \,\cite{SolGo2015} within the context of an arbitrary GUT, and also the result of \cite{LBS2014} in the different context of assuming a Gibbons-Hawking initial temperature. This finding, in its various versions, might provide a new solution to the entropy/horizon problems\,\cite{KolbTurner}.

\section{The current Universe}

Finally, in the matter-dominated epoch: $H^{2+n}\ll H^2$ in
(\ref{LambdaH2Hn}). The field equations can now be solved anew (with
$\omega_m=0$) using the neglected terms during the de Sitter period.
The energy densities are easily found:
\begin{equation}\label{eq:currentU}
\rho_{m}(a)=\rho_{m 0} a^{-3(1-\nu)}\,,\ \ \ \ \ \ \rho_{\Lambda}(a)=\rho_{\Lambda 0}+\frac{\nu\,\rho_{m
0}}{1-\nu}\left[a^{-3(1-\nu)}-1\right]\,,
\end{equation}
with $8\pi\,G\rho_{\Lambda
0}=3c_0+3\nu H_0^2$. They clearly follow the $\Lambda$CDM behavior
provided the condition $|\nu|\ll1$ holds\,\cite{PhenoVacuum1,PhenoVacuum2}, as theoretically
expected\,\cite{Fossil07}. Hints of dynamical vacuum
energy can indeed be detected even for $|\nu|$ as small as $10^{-3}$, see Ref.\cite{SolaGomCruz2015}.

It is also remarkable that the small transfer of energy between vacuum and matter embodied in (\ref{eq:currentU}), which is parameterized by $|\nu|\ll1$, can be interpreted (see \cite{FritzschSola1,Sola2014,FritzschSola2} and references therein) as a time variation of the particle masses (both from baryons and dark matter) and the fundamental ``constants'' of Nature. It is therefore tempting to propose (according to the aforementioned references) that such dynamical vacuum framework may offer a possible explanation within General Relativity for the numerous hints suggesting a small cosmic drift of their values -- see the recent \,\cite{Flambaum2015} and references therein. See also Ref. \cite{Sola2015Preface} for a summarized introduction to this fascinating subject.

\section{Conclusions}

In summary, the cosmological constant problem seems to be connected with the terms that also appear when one naively computes the vacuum energy in flat spacetime, whereas the (much smaller) contributions related with the dynamical curvature of the expanding Universe can be described in an effective way by quantum effects that follow a renormalization group equation driven by the Hubble flow.  As a result the Dark Energy that we measure today from the accelerated phase of our Cosmos should ultimately be the effect of the dynamical vacuum energy of the expanding background. After renormalizing away the terms that are in common with flat spacetime, we are left with a vacuum energy density and entropy which nicely resonate with all the main traits of the cosmic history, and with a dynamical tail that may be the ``smoking gun'' of this mechanism.  It all happens as though some of the most puzzling mysteries of our present may have indeed profound roots in the distant past -- those early times when our Universe encoded the fundamental seeds of its entire future evolution.


\vspace{1cm}

{\bf Acknowledgments}

\noindent I have been partially supported by FPA2013-46570 (MICINN), Consolider grant
CSD2007-00042 (CPAN), MDM-2014-0369 (ICCUB) and by 2014-SGR-104 (Generalitat de Catalunya). It is my pleasure to thank  S. Basilakos, H. Fritzsch, A. G\'omez-Valent, J.A.S. Lima and N. E. Mavromatos  for collaboration in previous works related with some of the matters presented here.


\begin{thebibliography}{99}


\bibitem{Weinberg89}S. Weinberg,
Rev. Mod. Phys. {\bf 61} (1989) 1

\bibitem{CCP1}  V. Sahni, A. Starobinsky, Int. J. of Mod. Phys. {A9} {(2000)} {373}.

\bibitem{CCP2} T. Padmanabhan, Phys. Rept.  {\bf 380} (2003)  235.

\bibitem{CCP3}  P.J. Peebles, B. Ratra, Rev. Mod. Phys. {\bf 75}
    (2003) 559.


\bibitem{BirrellDavis} N.D. Birrell and P.C.W. Davies,
    \textit{Quantum Fields in Curved Space} (Cambridge U. Press, 1982).

\bibitem{JSPRev2013}  J. Sol\`a,
J. Phys. Conf. Ser. {\bf 453} (2013)  012015 [arXiv:1306.1527];  AIP
Conf.Proc. {\bf 1606} (2014) 19 [arXiv:1402.7049].

\bibitem{ShapSol02} I.~L. Shapiro and J.~Sol{\`a},  JHEP {\bf 0202} (2002) 006; Phys.Lett. {\bf B682} (2009) 105.

\bibitem{Fossil07} J.~Sol{\`a},  J. Phys. Math. Theor. A {\bf 41} (2008) 164066.

\bibitem{BasMavroSol2015} S. Basilakos, N. E. Mavromatos, and J.~Sol{\`a}, 
    e-Print: arXiv:1505.04434.

\bibitem {Markkanen2015} T. Markkannen, Phys.Rev. D{\bf 91} (2015) 124011.


\bibitem{LBS2013} J.~A.~S.~Lima, S.~Basilakos and J.~Sol\`a,
  Mon.\ Not.\ Roy.\ Astron.\ Soc.\  {\bf 431} (2013) 923.

\bibitem{LiddleLyth} A.R. Liddle, D. H. Lyth, \textit{Cosmological Inflation and Large-Scale Structure} (Cambridge U. Press, New York, 2000); \textit{The Primordial Density Perturbation} (Cambridge U. Press, New York, 2009).

\bibitem{KolbTurner} E. W. Kolb, M.S. Turner, \textit{The Early Universe} (Addison-Wesley, MA, 1990).

\bibitem{SolGo2015}
J. Sol\`{a}, A. G\'omez-Valent, Int. J. Mod. Phys. {\bf D24} (2015) 1541003.

\bibitem{LBS2014} J.A.S. Lima, S, Basilakos, J. Sol\`a,
 Gen. Rel. Grav. {\bf 47} (2015)  40.

\bibitem{PhenoVacuum1} A. G\'omez-Valent, J. Sol\`{a}, S. Basilakos,
    JCAP {\bf 1501} (2015) 004.

\bibitem{PhenoVacuum2} A. G\'omez-Valent, and  J. Sol\`{a}, Mon. Not. Roy. Astron. Soc. {\bf 448} (2015) 2810.

\bibitem{SolaGomCruz2015}  	
J. Sol\`a, A. G\'omez-Valent, J. de Cruz P\'erez, Astrophys.J. {\bf 811} (2015) L14 [e-Print: arXiv:1506.05793];  A. G\'omez-Valent, E. Karimkhani, J. Sol\`a,  	
e-Print: arXiv:1509.03298.


\bibitem{FritzschSola1} H. Fritzsch, J. Sol\`a, Class. Quant. Grav.
    {\bf 29} (2012) 215002;  Adv.High Energy Phys. {\bf 2014} (2014) 361587.

\bibitem{Sola2014}    J. Sol\`a, Int. J.  Mod.  Phys. {\bf A29} (2014) 1444016.

 	
\bibitem{FritzschSola2} H. Fritzsch, J. Sol\`a,  Mod.Phys.Lett. A{\bf 30} (2015) 1540034.

\bibitem{Flambaum2015} 	
Y.V. Stadnik, V.V. Flambaum, Phys. Rev. Lett. {\bf 114} (2015) 161301; 	
e-Print: arXiv:1503.08540.

\bibitem{Sola2015Preface} 	J. Sol\`a, Mod. Phys. Lett. A {\bf 30} (2015) 1502004 [e-Print: arXiv:1507.02229].


\end{thebibliography}
\end{document}